\documentclass[twocolumn]{article}
\usepackage{geometry}
\usepackage{mathtools}
\usepackage{graphicx}
\usepackage{color}
\usepackage{upgreek}
\usepackage{float}
\usepackage{txfonts}
\usepackage{longtable}
\usepackage{caption}
\usepackage[comma,super,sort&compress]{natbib}
\usepackage[colorlinks = true,
            linkcolor = blue,
            urlcolor  = blue,
            citecolor = blue,
            anchorcolor = blue]{hyperref}

\newcommand{\apj}{Astrophys. J.} 
\newcommand{\apjs}{Astrophys. J. Suppl. Ser.} 
\newcommand{\aj}{AJ}                                 
\newcommand{\araa}{Annu. Rev. Astron. Astrophys.} 
\newcommand{\mnras}{Mon. Not. R. Astron. Soc.} 
\newcommand{\nat}{Nature}

\newcommand{\aap}{Astron. Astrophys.} 
                                
\newcommand{\aapr}{Astron. \& Astrophys.\ Rev.} 
\newcommand{\physrep}{Phys. Rep.} 
\newcommand{\apjl}{Astrophys. J. Lett.} 

\newcommand{\pasp}{Publ. Astron. Soc. Pac.} 
\newcommand{\pasa}{Pubs. Astron. Soc. Australia}

\def\frb{FRB\,20180916B}
\def\psr{PSR\,B2111$+$46}

\makeatother

\begin{document}

\title{Highly polarised microstructure from the repeating \frb}

\author{K.~Nimmo$\mathrm{^{*}}$\\
        ASTRON, Netherlands Institute for Radio Astronomy,\\ Oude Hoogeveensedijk 4, 7991 PD Dwingeloo, The Netherlands \\
        Anton Pannekoek Institute for Astronomy, University of Amsterdam, \\Science Park 904, 1098 XH Amsterdam, The Netherlands\\
         \texttt{k.nimmo@uva.nl} 
         \and
        J.~W.~T.~Hessels \\
         ASTRON, Netherlands Institute for Radio Astronomy,\\ Oude Hoogeveensedijk 4, 7991 PD Dwingeloo, The Netherlands \\
        Anton Pannekoek Institute for Astronomy, University of Amsterdam, \\Science Park 904, 1098 XH Amsterdam, The Netherlands
        \and
        A.~Keimpema \\
        Joint Institute for VLBI ERIC (JIVE), \\Oude Hoogeveensedijk 4, 7991 PD Dwingeloo, The Netherlands
        \and
        A.~M.~Archibald \\
        School of Mathematics, Physics, and Statistics, University of Newcastle,\\ Newcatle upon Tyne, NE1 7RU, United Kingdom
        \and
        J.~M.~Cordes \\
        Department of Astronomy and Cornell Center for Astrophysics and Planetary Science, \\Cornell University, Ithaca, NY 14853 USA
        \and
        R.~Karuppusamy \\
        Max-Planck-Institut f\"ur Radioastronomie, \\Auf dem H\"ugel 69, D-53121 Bonn, Germany
        \and
        F.~Kirsten \\
        Department of Space, Earth and Environment, Chalmers University of Technology, \\ Onsala Space Observatory, 439 92, Onsala, Sweden
        \and
        D.~Z.~Li \\
        Department of Physics, University of Toronto,\\ 60 St. George Street, Toronto, ON M5S 1A7, Canada\\
        Canadian Institute for Theoretical Astrophysics, \\60 St. George Street, Toronto, ON M5S 3H8, Canada
        \and 
        B.~Marcote \\
        Joint Institute for VLBI ERIC (JIVE), \\Oude Hoogeveensedijk 4, 7991 PD Dwingeloo, The Netherlands
        \and
        Z.~Paragi \\
        Joint Institute for VLBI ERIC (JIVE), \\Oude Hoogeveensedijk 4, 7991 PD Dwingeloo, The Netherlands \\
         \\
         \\
        }
          
\maketitle

\section*{Abstract} 
   Fast radio bursts (FRBs) are bright, coherent, short-duration radio transients of as-yet unknown extragalactic origin. FRBs exhibit a wide variety of spectral, temporal and polarimetric properties, which can unveil clues into their emission physics and propagation effects in the local medium. Here we present the high-time-resolution (down to 1\,$\upmu$s) polarimetric properties of four 1.7-GHz bursts from the repeating \frb, which were detected in voltage data during observations with the European VLBI Network (EVN). We observe a range of emission timescales spanning three orders of magnitude, with the shortest component width reaching 3--4\,$\upmu$s (below which we are limited by scattering). This is the shortest timescale measured in any FRB, to date. We demonstrate that all four bursts are highly linearly polarised ($\gtrsim 80\%$), show no evidence for significant circular polarisation ($\lesssim 15$\%), and exhibit a constant polarisation position angle (PPA) during and {\it between} bursts. On short timescales ($\lesssim 100$\,$\upmu$s), however, there appear to be subtle (few degree) PPA variations across the burst profiles. These observational results are most naturally explained in an FRB model where the emission is magnetospheric in origin, as opposed to models where the emission originates at larger distances in a relativistic shock.

\section{Introduction} 

Many FRBs\cite{Petroff2019,Cordes_Chatt_Rev} show complex burst morphology and, to date, both repeaters and apparent non-repeaters have shown temporal structure, `sub-bursts', as short as 20--30\,$\upmu$s \citep{Farah2018,Michilli2018,Cho2020}.  Probing even shorter, $1$\,$\mathrm{\upmu s}$, timescales is a powerful way to constrain emission models because of the limits that such temporal structures place on the instantaneous size of the emitting region: $1$\,$\mathrm{\upmu s}$ corresponds to 300\,m, though special relativistic effects cause the light-travel size to be much smaller than the actual size. 

Voltage data allow access to such timescales, but several practical challenges remain: e.g., scattering due to multi-path propagation can limit the effective time resolution (especially at low radio frequencies); the signal-to-noise (S/N) on short timescales may be too low; there may be limitations on the precision with which the dispersion measure (DM) can be determined, such that it is impossible to ensure that the DM smearing is less than the time resolution; and, if the bursts are composed of a forest of closely spaced (sub-)$\mathrm{\upmu s}$ sub-bursts, then confusion may limit our ability to identify individual structures.  For example, in the case of FRB\,20181112A \citep{Cho2020}, the effective time resolution is limited by scattering, despite having voltage data. 

Ultra-high-time-resolution studies are even more powerful if they include full polarisation information.  In general, FRBs show a wide variety of polarimetric properties. FRBs have been observed to exhibit linear polarisation fractions from 0$\%$ to 100$\%$ (e.g. \cite{Masui2015,Michilli2018,Caleb2018,Day2020}). Some FRBs show significant circular polarisation\cite{Petroff2015}, though most so far show very little\cite{Ravi2016,Caleb2018}. Some sources show a flat PPA across the burst profile \citep{Michilli2018,R3_disc,Day2020}, whereas others show a PPA variation \cite{Masui2015,Luo2020}.

The polarisation properties of FRB\,20121102A, the first discovered repeating FRB \citep{Spitler2014,Spitler2016}, were measured at $4$--$8$\,GHz with Arecibo\cite{Michilli2018} and the Green Bank Telescope (GBT) \cite{, Gajjar2018}. FRB\,20121102A bursts are approximately 100$\%$ linearly polarised, and show no sign of circular polarisation. There is evidence of a linear polarisation decrease towards lower frequencies but whether this is intrinsic or reflects a propagation effect is currently unclear (Plavin et al., in prep.). Additionally, for FRB\,20121102A, the PPA is flat across the burst duration, and the absolute value of PPA is approximately equal between bursts  \citep{Michilli2018}. The rotation measure (RM) of FRB\,20121102A was found to be very large (10$^{5}\,\mathrm{rad\,m^{-2}}$), and highly variable (variation of approximately 10$\%$ over $7$ months), implying an extreme and dynamic local magneto-ionic environment \citep{Michilli2018,hilmarsson2020}. 

FRB\,20190711A has recently been shown to repeat \citep{Day2020,Kumar2020}. As with FRB\,20121102A, the bursts show high fractional linear polarisation (approximately 80--100\%), no significant circular polarisation, and a flat PPA across the burst profile. The RM of FRB\,20190711A was found to be $9\pm 1\,\mathrm{rad\,m^{-2}}$, four orders of magnitude smaller than in the case of FRB\,20121102A.

\citet{Luo2020} report the discovery of repeat bursts from FRB\,20180301A, with the Five-hundred-meter Aperture Spherical radio Telescope (FAST), and present the polarisation properties of $7$ bursts from this source at 1.25\,GHz. Similar to other repeating FRBs, these bursts show no evidence of circular polarisation. However, contrary to the properties of previously studied repeaters, the linear polarisation fractions are measured to be lower (approximately $30$--$80\%$), and some bursts show a swing in the PPA across the burst profile. The RM measured for this source varies from $521.5\,\mathrm{rad\ m^{-2}}$ to $564.4\,\mathrm{rad\ m^{-2}}$ between bursts. 

The only other repeating FRB with published polarisation information from multiple bursts is \frb\ \citep{R3_disc, Chawla2020}.  The RM measured for \frb\ (about $ -115\,\mathrm{rad\ m^{-2}}$) is three orders of magnitude smaller than what is seen for FRB\,20121102A \cite{R3_disc}. \citet{R3_disc} discuss the capture of CHIME/FRB voltage data during one burst from \frb\ in the frequency range $400$--$800\,\mathrm{MHz}$. With this data, measurements of the polarisation fractions and RM were first possible. This burst exhibits approximately 100\% linear polarisation, and shows no evidence for circular polarisation. The PPA appears to be flat over the burst duration.  The polarimetric properties of four bursts from \frb, detected at $300$--$400\,\mathrm{MHz}$ using GBT, are consistent with the original discovery \cite{Chawla2020} (approximately 100\% linear, flat PPA, comparable RM) --- supporting the idea that repeating FRBs have consistent polarisation properties between bursts from the same source (over a range of radio frequencies), and that the phenomenology is similar for repeating FRB sources in general. The polarisation properties of \frb, measured with LOFAR from $110$--$188$\,MHz, are similar to those at higher frequencies, but also show significant depolarisation down to lowest observed frequencies\cite{pleunis2020}.  This is likely the result of scattering, though investigation into whether this is due to Faraday conversion\cite{gruzinov2019,vedantham2019} is underway\cite{pleunis2020}.

\frb's polarimetric properties have not previously been investigated at radio frequencies $> 1$\,GHz, where most FRB\,20121102A studies have been conducted, and previous studies have been limited by the temporal resolution of the data. It is unclear if the polarimetric properties seen from \frb\ at $110$--$800$\,MHz persist at higher radio frequencies, and the shortest temporal structure reported to date is $60$\,$\mathrm{\upmu s}$ \citep{Marcote2020}. Also, at higher frequencies, measurements of the PPA are less affected by RM variations, and thus easier to study. 

Here we present an analysis of four 1.7-GHz \frb\ bursts whose spectro-temporal properties were previously presented \cite{Marcote2020}.  In the present work, we provide a higher-time-resolution, full polarimetric analysis of this sample.  Throughout this paper we use the nomenclature B$n$ for the bursts, following \citet{Marcote2020}. Additionally, we introduce the nomenclature B4-sb$n$ for the three clear sub-bursts in burst B4. In \S2 we describe the data, in \S3 we present the high-time-resolution and polarimetry results, and thereafter discuss the consequences for our understanding of repeating FRBs in \S4. 

\section{The data}
The data were acquired as part of an EVN campaign on 2019 June 19 (experiment code: EM135C) at a central radio frequency of $1.7$\,GHz. Using SFXC \citep{keimpema2015}, we created two data products. Firstly, we converted the voltage data from the $100$-m Effelsberg telescope into full-polarisation (circular basis) filterbank data with time and frequency resolutions of $1\,\mathrm{\upmu s}$ and $0.5\,\mathrm{MHz}$, respectively. Secondly, we produced full-polarisation filterbank data with time and frequency resolutions of $16\,\mathrm{\upmu s}$ and $62.5\,\mathrm{kHz}$, respectively. In this process, the data were coherently dedispersed to a dispersion measure (DM) of $348.76\,\mathrm{pc\ cm^{-3}}$, which is the best-fit DM that maximises S/N for burst B4 at 16\,$\upmu$s time resolution \citep{Marcote2020}. Using PSRCHIVE \citep{vanStraten_psrchive_2011}, we created archive files containing each burst from the filterbank files at the native time and frequency resolution. The archive files are incoherently dedispersed to a refined DM using the 1\,$\upmu$s data (see Methods). We manually mask frequency channels that are contaminated by radio frequency interference (RFI), and artefacts at the sub-band edges. 

\section{Results}
\subsection{High time resolution}
In our previous spectral and temporal characterisation of these bursts \citep{Marcote2020}, we found that B3 and B4 show several sub-bursts with widths ranging from $60$--$700$\,$\upmu$s. Here we probe order-of-magnitude shorter timescales by studying the burst temporal properties at even higher time resolution. 

\begin{figure*}
\resizebox{\hsize}{!}
        {\includegraphics[trim=0cm 0.5cm 0cm 1.0cm, clip=true,width=\textwidth,height=200mm]{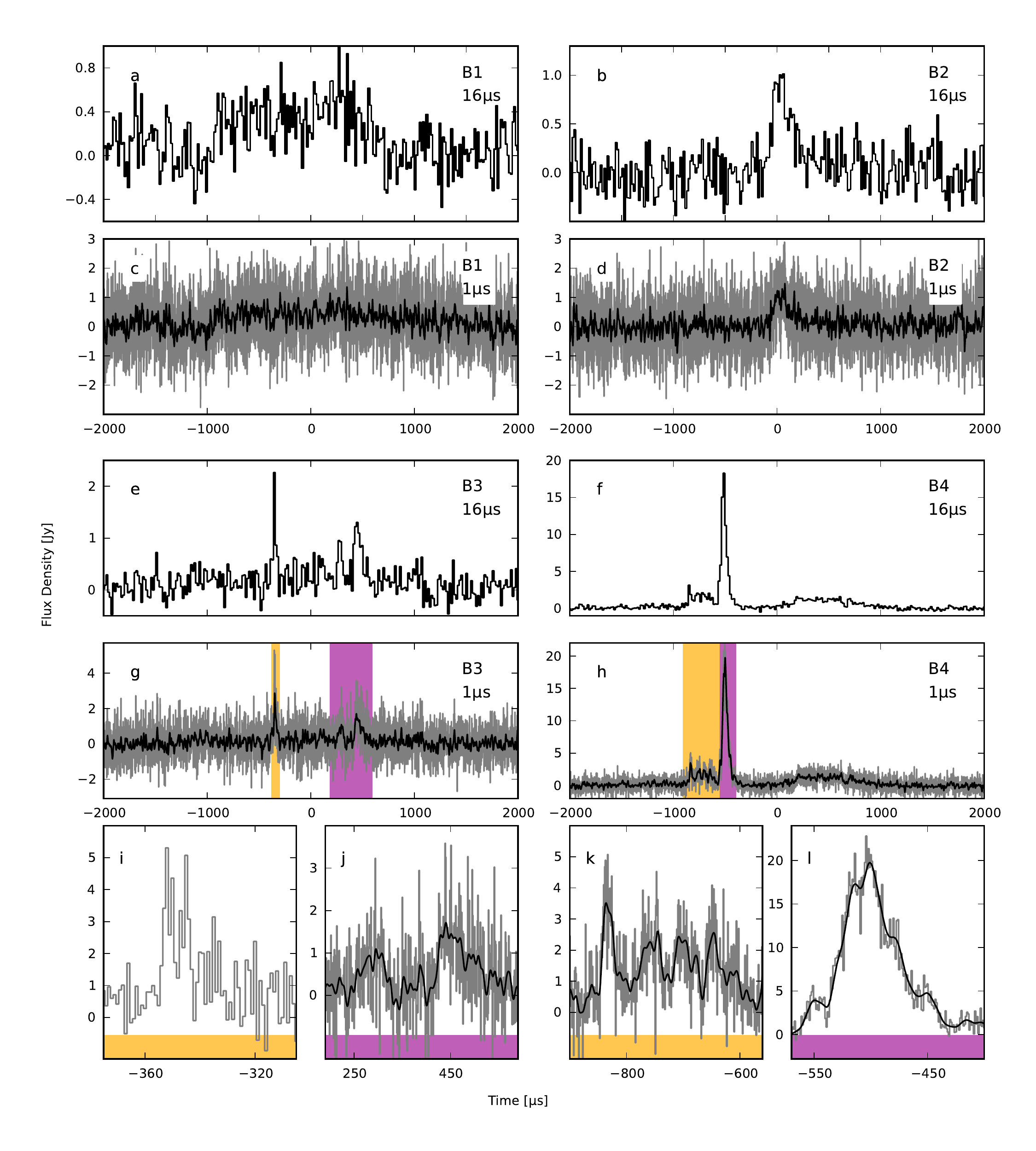}}
  \caption{Burst profiles at 16\,$\upmu$s and 1\,$\upmu$s time resolution for four 1.7\,GHz bursts from \frb. The burst name (B1--B4) and time resolution used for plotting is shown in the top right of each panel. Bursts B3 and B4 show complex temporal structure. Panels i and j are zoomed-in 1\,$\upmu$s resolution data around the B3 burst components highlighted by the orange and purple bars in panel g, respectively. Similarly for burst B4, panels k and l are zoomed-in 1\,$\upmu$s resolution  data around the B4 burst components highlighted by the orange and purple bars in panel h, respectively. Overplotted on panels j, k and l is a smoothed profile using a Blackman window function with a window length of 19 bins. }
     \label{fig:burstsplot_time}
\end{figure*}

In Figure\,\ref{fig:burstsplot_time}, we present the four \frb\ burst profiles at both $16$\,$\upmu$s and $1$\,$\upmu$s resolution. In the case of bursts B1 and B2, the increase in time resolution does not reveal any shorter-timescale structure, and the burst widths are consistent with the widths measured previously\cite{Marcote2020} (1.86\,$\pm$0.13\,ms and 0.24\,$\pm$\,0.02\,ms for B1 and B2, respectively). Bursts B3 and B4 show clear structure in the 16\,$\upmu$s resolution data. By increasing the time resolution to 1\,$\upmu$s, we see clear 10--20\,$\upmu$s structure in burst B4 (panel k, Figure\,\ref{fig:burstsplot_time}), and the bright component of B4 does not appear to be a simple Gaussian envelope; instead, it exhibits 10--20\,$\upmu$s fluctuations on top of the broader envelope (panel l, Figure\,\ref{fig:burstsplot_time}). B3 exhibits 50--100\,$\upmu$s components (panel j, Figure\,\ref{fig:burstsplot_time}), and even a component that is only a few $\upmu$s wide (panel i, Figure\,\ref{fig:burstsplot_time}). We note that the two 3--4\,$\upmu$s components shown in panel i, Figure\,\ref{fig:burstsplot_time}, are detected across our 128\,MHz band. The 3--4\,$\upmu$s shots of emission are consistent with the estimated Galactic scattering time (2.7\,$\upmu$s) \cite{Marcote2020} for \frb.

\begin{figure*}
\resizebox{\hsize}{!}
        {\includegraphics[trim=0cm 1.5cm 0cm 3.05cm, clip=true,height=220mm]{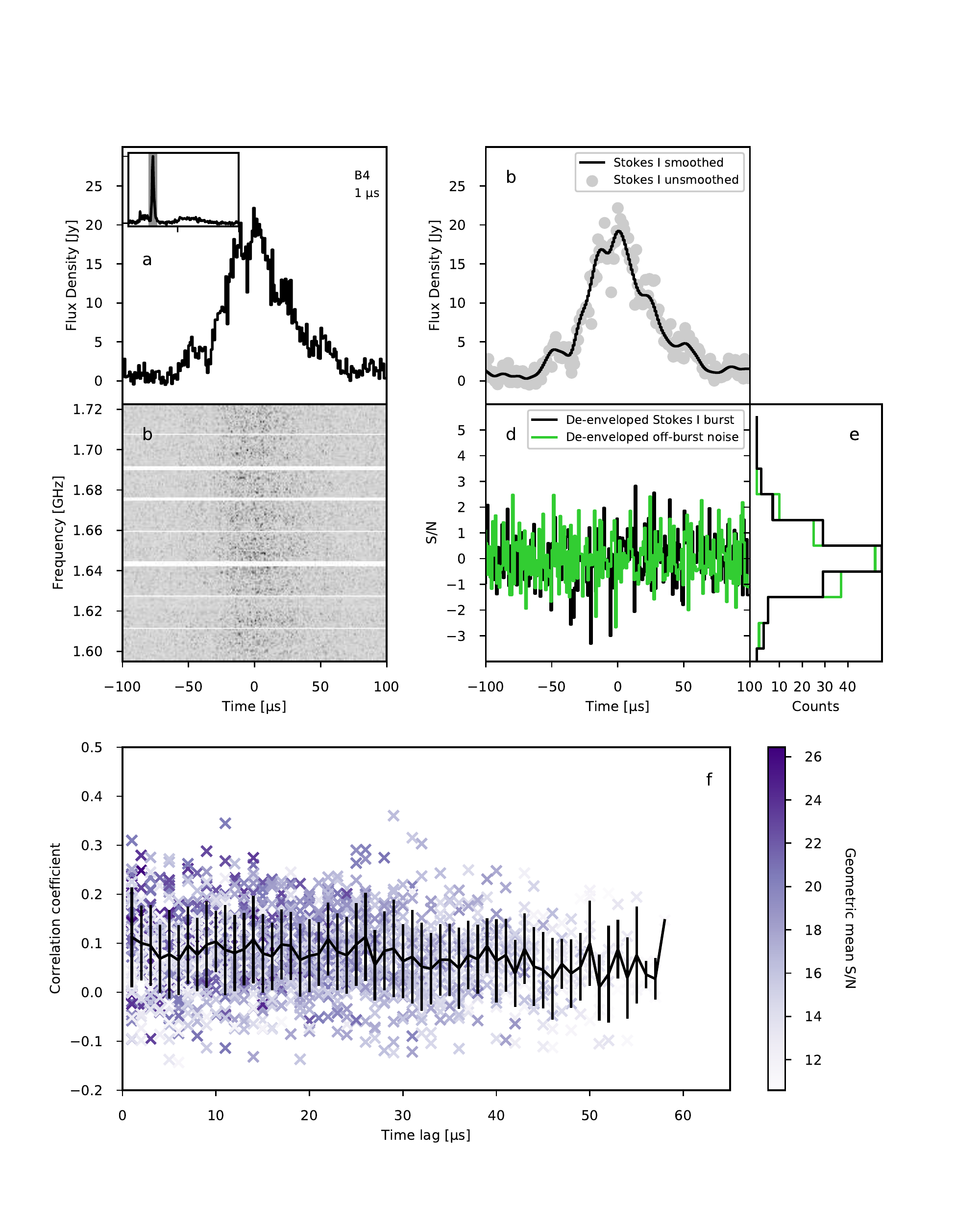}}
  \caption{Stokes~I burst profile of the bright component of burst B4 (B4-sb2) sampled at 1\,$\upmu$s resolution (panel a). The inset of panel a shows the profile of B4 at 16\,$\upmu$s resolution, with the shaded region indicating the extent of the profile shown in the panel. The dynamic spectrum is shown in panel b. Panel c shows the profile again, overplotted with a smoothed profile using a Blackman window function with a window length of 19 bins. The burst profile was ``de-enveloped" by dividing out the smoothed profile, and the residuals are shown in black in panel d. Also shown in panel d is the de-enveloped off-burst time series in green, for comparison. In panel e, we show the histogram of the on-burst and off-burst residuals. Panel f shows the correlation coefficient of the spectra of bins with S/N$>10$ in the main peak of B4, as a function of time between the spectra in $\upmu$s. The colour bar represents the geometric mean S/N of the two time bins that have been correlated, with the darker purple representing a higher S/N. Overplotted in black is the weighted mean and standard deviation per time lag.}
     \label{fig:amp_mod_noise}
\end{figure*}

In addition to the 10--20\,$\upmu$s structure in the profile of B4-sb2 (panel l, Figure\,\ref{fig:burstsplot_time}), there also appear to be narrower spikes on top of the burst envelope. We fit the burst envelope and remove it from the data, in order to test the statistical significance of any outliers (see Methods). We find no significant outliers in this burst, implying that the narrow features are consistent with amplitude-modulated noise (Figure\,\ref{fig:amp_mod_noise}). This is further supported by our measurement of a S/N-weighted correlation coefficient $<$\,0.2 (panel f; Figure\,\ref{fig:amp_mod_noise}), where we would expect a correlation coefficient of 1/3 if each time bin is perfectly correlated\cite{cordes2004} (and the scattering time is greater than the separation of the bursts). Due to the large scatter, we cannot distinguish between a constant or slightly decreasing correlation coefficient as a function of lag. Thus, based on the data in hand, we can rule out that burst B4 is comprised of a few well-separated bright (sub) $\upmu$s shots of emission.  

By eye, B4-sb1 (panel k, Figure\,\ref{fig:burstsplot_time}) appears to fluctuate quasi-periodically. There are other examples of FRBs showing this diffraction-pattern-like temporal behaviour\cite{Hessels2019}, which potentially can be explained by self-modulation breaking the burst into `pancakes' \cite{sobacchi2021}, or by plasma lensing \cite{cordes2017}. To test this, we computed the autocorrelation function (ACF; using Equation~1 in \citet{Marcote2020}, but here considering time lag instead of frequency lag), and the power spectrum (Extended Data Figure\,\ref{fig:ACFPS}). For details of the power spectrum modelling and statistics, see Methods. We find no statistically significant evidence for quasi-periodic emission in the power spectrum of B4-sb1 or B3. We conclude that bursts B4-sb1 and B3 are consistent with red noise, with a power-law index of $\alpha =$\,1.58\,$\pm$\,0.02 and  $\alpha =$\,1.38\,$\pm$\,0.01, respectively. Typically, magnetar X-ray bursts show steeper red noise spectra\cite{huppenkothen2013} ($\alpha\approx$\,2--5). 

\subsection{Polarimetry}
As described in Methods, the polarimetric data were calibrated using a test pulsar (\psr).

For the highest-S/N burst in our sample, B4, we measure the RM to be $-104\,\pm\,20$\,rad\,m$^{-2}$ (see Methods), where the large fractional error arises due to covariances between the Q-U fit parameters and the instrumental delay between the two polarisation channels, which is not independently constrained. We find the RM to be consistent with the previously measured RM values for \frb\ \citep{R3_disc,Chawla2020}.

\begin{figure*}
\resizebox{\hsize}{!}
        {\includegraphics[trim=0cm 0.6cm 0cm 0.6cm, clip=true]{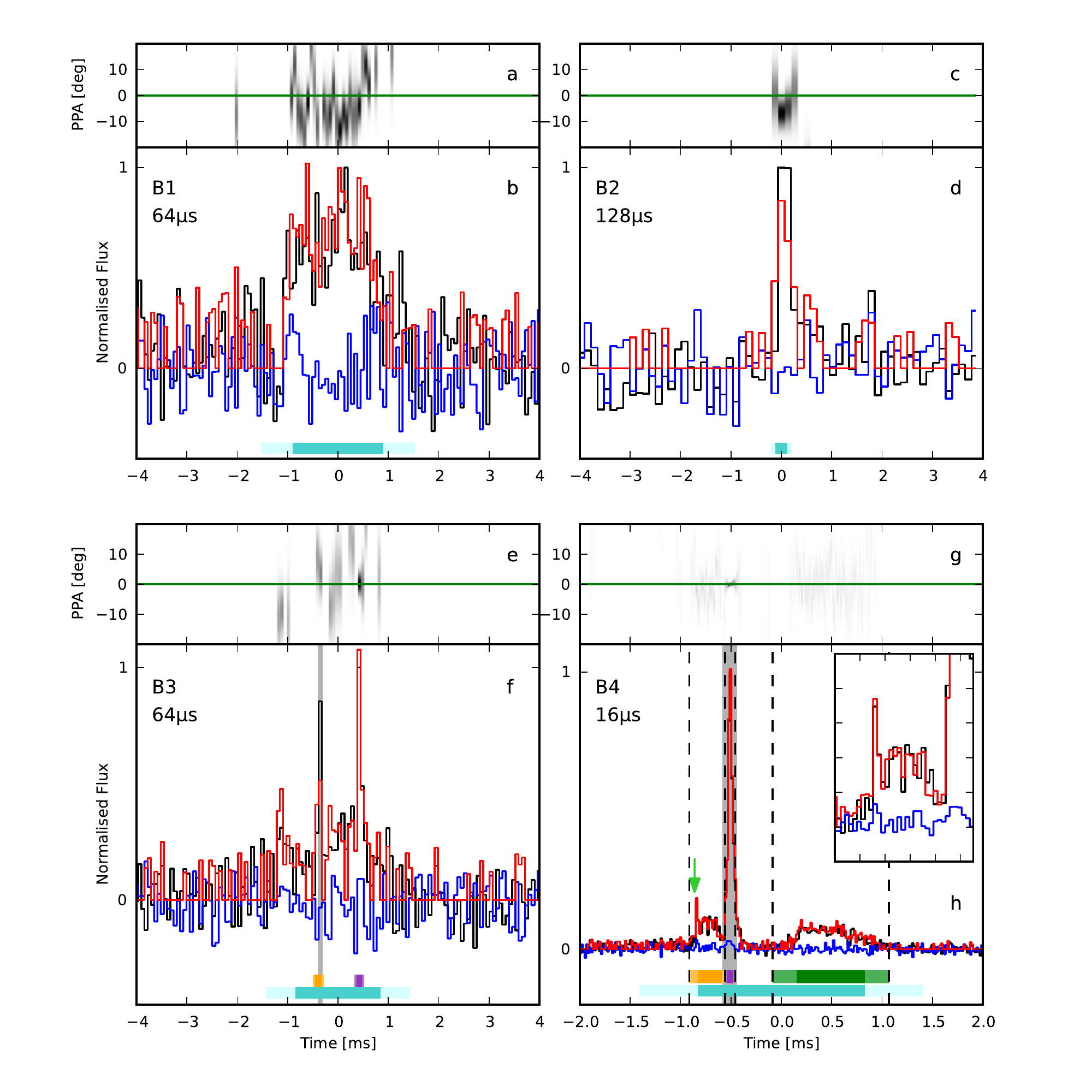}}
  \caption{Polarimetric profiles (lower panels of each sub-figure) and PPAs (top panels of each sub-figure) for the four bursts from \frb\ discovered during an EVN campaign on 2019 June 19 \citep{Marcote2020}. In the bottom panels, the total intensity (Stokes~I) profile is shown in black, the unbiased linear polarisation (Equation \ref{eq:Lunbias}) is shown in red, and circular polarisation (Stokes~V) is shown in blue. For B1, B2 and B3, we plot $8\, \mathrm{ms}$ around the burst, and for B4 we plot $4\,\mathrm{ms}$. The time is referenced to the mean of the Gaussian fit to the burst envelope discussed in \citet{Marcote2020}. The inset in panel h shows a zoom-in on the profile at the leading edge, highlighting a narrow, approximately 30\,$\mathrm{\upmu s}$, spike, also highlighted by the green arrow. The top left of each lower panel shows the burst name, B$n$, used to define the bursts in this work (ordered according to their arrival time), and the time resolution used for plotting. Also shown in the lower panels are the Gaussian full-width at half-maximum (FWHM) of each burst illustrated by the dark cyan bar. The light cyan bar represents the 2-$\sigma$ region. Bursts B3 and B4 show multiple sub-bursts indicated by the orange, purple and green bars in panels f and h (the FWHM is shown in the dark colour, and the 2-$\sigma$ region shown in the lighter colour). For burst B4, we also show dotted lines indicating the extent of the three sub-bursts: B4-sb1 (orange), B4-sb2 (purple) and B4-sb3 (green). The top panel shows the PPA, defined as $\mathrm{PPA}=0.5\tan^{-1}{({\rm U}/{\rm Q})}$. The greyscale represents the probability distribution of the PPA following \citet{everett2001}, the darker shading representing higher polarised S/N. The PPA has been shifted by the best-fit flat PPA of the four bursts, weighted by their unbiased linear polarisation S/N.  Thus, the weighted PPA of the four bursts is set to zero, as is illustrated by the green line. The shaded region of the profile of B3 and B4 highlights the timescale plotted in Figure\,\ref{fig:high_time_pol} for each burst. } 
     \label{fig:burstsplot}
\end{figure*}

In Figure~\ref{fig:burstsplot}, we show the Faraday-derotated profiles for the four \frb\ bursts. We use the {\tt rmfit}-determined RM for B4 (see Methods) to correct all four bursts, since B4 has the highest S/N, and we assume the RM does not change between bursts separated by approximately 4$\,\mathrm{hours}$. We plot the unbiased linear polarisation, $L_{\rm unbias}$, profile in red, following \citet{everett2001}, where
\begin{equation} \label{eq:Lunbias} L_{\rm unbias}=\begin{cases}
    \sigma_{I}\sqrt{\left(\frac{L_{\rm meas}}{\sigma_{I}}\right)^2-1}, & {\text{if} \frac{L_{\rm meas}}{\sigma_{I}}\ge 1.57}\\
    0, & \text{otherwise}
  \end{cases}  \end{equation} 
where $L_{\rm meas} = \sqrt{Q^2+U^2}$, and $\sigma_{I}$ is the standard deviation in the off-burst Stokes~I data. 

The PPA, corrected for parallactic angle variations (see Methods), is shown in the top panel of each sub-figure in Figure~\ref{fig:burstsplot}. We performed a least-squares fit of a horizontal line to the PPA of the four bursts together, weighted by their $1$-$\sigma$ errors. We note that for all of the PPA fits, we consider only additive noise in the determination of the variance. The weighted $\chi^{2}$-statistic for this global fit is $175.5$, with $125$ degrees of freedom. We have shifted the absolute value of the PPA by this best-fit value, $-89.2^{\circ}$. We note that, due to imperfect calibration, this value is not the absolute PPA and should not be used for comparison with bursts from other studies of \frb. We performed individual least-square fits for each burst, the results of which are reported in Extended Data Table\,\ref{tab:pol_properties}. For the above fits, we only included PPAs within the Gaussian-fit 2-$\sigma$ temporal width region (illustrated by the light cyan bars shown in Figure~\ref{fig:burstsplot}) that also satisfied $L_{\rm unbias}/\sigma_{I} >3$. We find that the PPAs of the four bursts are consistent with being constant across the burst duration. We do, however, see a hint of PPA variation between burst components (which is most evident in B1 and B3).

\begin{figure*}
\resizebox{\hsize}{!}
        {\includegraphics[trim=0cm 0.6cm 0cm 0.6cm, clip=true]{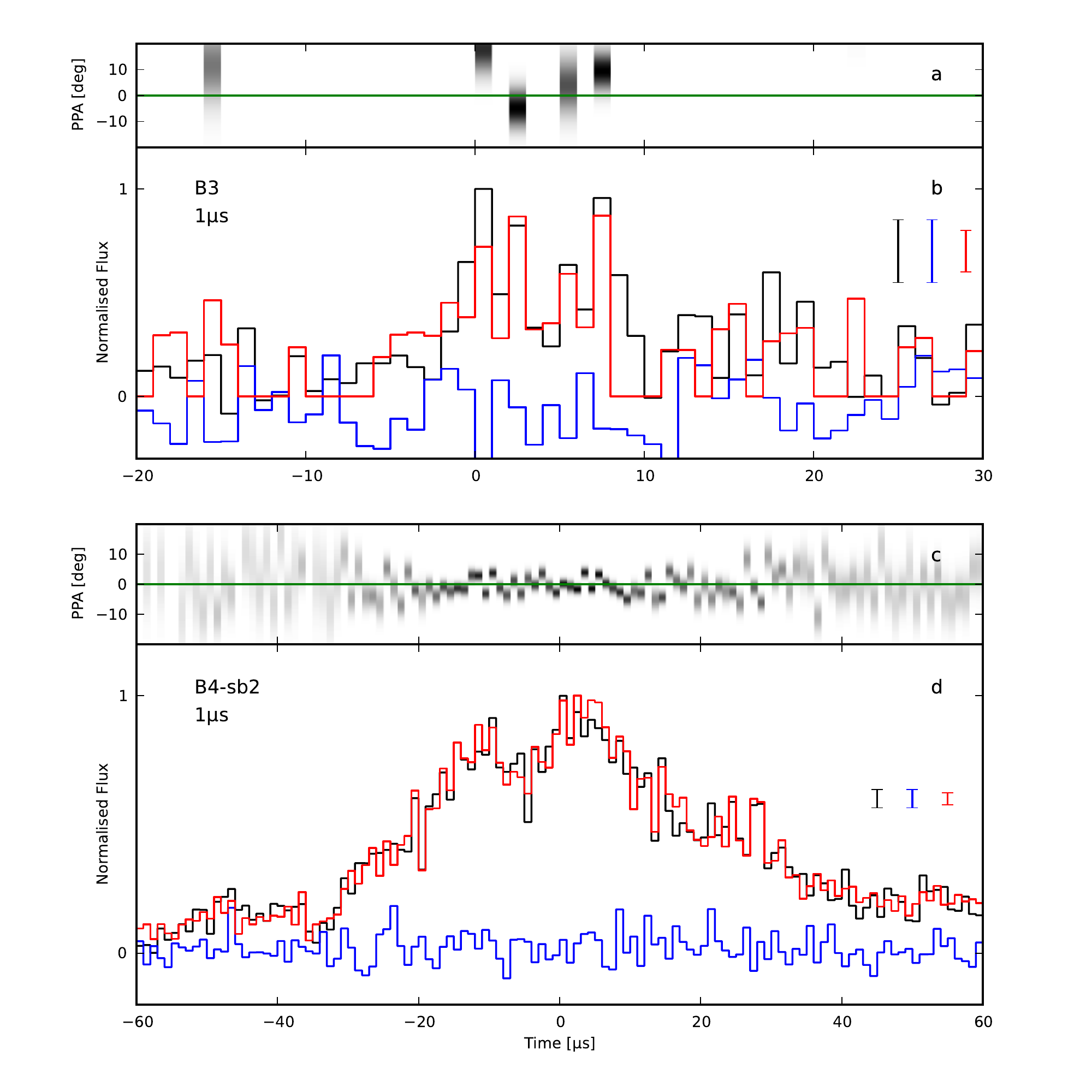}}
  \caption{Polarisation profiles (panels b and d) and polarisation position angle (PPA; panels a and c) of burst components of B3 (top figure) and B4 (bottom figure), plotted at 1\,$\upmu$s time resolution. The shaded regions of the burst profiles shown in Figure\,\ref{fig:burstsplot_time} highlight the timescales plotted here. On the right of the bottom panels, the off burst standard deviation is shown. As is also done in Figure\,\ref{fig:burstsplot}, the PPA has been shifted by the best-fit flat PPA of the four bursts weighted by their S/N.  The weighted PPA of the four bursts is thus set to zero, as illustrated by the green line.}
     \label{fig:high_time_pol}
\end{figure*}

We find that all four bursts are highly linearly polarised ($>80\%$), and show no evidence for circular polarisation ($<15\%$; Extended Data Table\,\ref{tab:pol_properties}). Additionally, the PPA of each burst is consistent with being flat across the burst duration, with the absolute PPA within approximately 7$^{\circ}$ between bursts. The second spike in the B3 profile (indicated by an orange bar in Figure\,\ref{fig:burstsplot}) appears to have a lower linear polarisation fraction than the rest of the burst. We see that the observed linear polarisation fraction increases when viewed in the higher-time-resolution 1\,$\upmu$s data (Figure\,\ref{fig:high_time_pol}). As we have shown in \S3.1, this component is actually only a few $\upmu$s wide (Figure\,\ref{fig:burstsplot_time}).  Therefore, the low polarisation fraction in the lower-time-resolution 16\,$\upmu$s data can be attributed to the fact that this component is not resolved. This highlights another importance of studying the burst properties at high time resolution.

In addition, we show the PPA and polarisation profile of B4-sb2 at 1\,$\upmu$s resolution in the lower sub-figure of Figure\,\ref{fig:high_time_pol}. At this resolution, there are small (approximately a few degrees) variations in the PPA across this bright burst component. To test the significance of these variations, we performed a weighted least-squares fit of a flat PPA to the 1\,$\upmu$s resolution PPAs across the bright burst component of B4. The measured reduced-$\chi^2$ of this fit is 3.7, compared with a reduced-$\chi^2 \approx$1 for the 16\,$\upmu$s resolution data. We conclude that the variations are significant.

\section{Discussion}\label{discussion} 
Neutron stars are prodigious generators of short-duration radio bursts, including canonical radio pulsar emission \citep{PSR_Handbook}, giant pulses \citep{Hankins2003,Hankins2007}, and radio pulses from magnetars \citep{Camilo2006}.  The Crab pulsar shows a variety of emission features at different radio frequencies \cite{Hankins2015}, each with their own characteristic spectro-tempo-polarimetric properties \citep{Hankins2016}.  \citet{Hessels2019} commented on the very similar phenomenology seen when comparing FRB\,20121102A with the high-frequency interpulses (HFIPs) produced by the Crab pulsar.  Like FRB\,20121102A and \frb, the Crab pulsar HFIPs typically show high (approximately 80--90\%) linear polarisation, weak (approximately 10--20\%) or undetectable circular polarisation, and non-varying PPA within and between bursts \citep{Hankins2016}. Since HFIPs are observed to be highly polarised, this implies that the emission region is spatially localised \cite{Hankins2016} (as opposed to coming from an extended region from the neutron star surface to the light cylinder, which would ultimately lead to depolarisation\cite{dyks2004}). Additionally, the flat PPAs between HFIPs suggest that the magnetic field direction is stable during each observational epoch. There have been examples of HFIPs, however, that deviate from this trend, either showing significant circular polarisation, weaker linear polarisation and/or a significant PPA variation across the burst profile \citep{Hankins2016}.

Comparing phenomenology with the Crab pulsar is tempting, but ignores the fact that FRB\,20121102A, \frb\ and other repeaters produce bursts that are orders of magnitude longer duration and higher luminosity.  Indeed, \citet{Lyutikov2017} argue that the established extragalactic distances of FRBs preclude rotational energy and support magnetic energy as the fundamental power source for the bursts.  Many FRB theories have invoked a magnetar as the central engine (see \citet{Platts2019} for a catalogue of FRB theories).  The recent discovery of an exceptionally bright (kJy--MJy) millisecond-duration radio burst from the Galactic magnetar SGR\,1935+2154 has added compelling evidence for such a scenario \citep{CHIME_SGR,Bochenek2020}.  In fact, SGR~1935+2154 has been observed to produce sporadic radio bursts spanning more than 7 orders-of-magnitude in fluence \citep{Kirsten2020}, though it is unclear whether these all arise from the same physical mechanism.

Magnetar FRB models come in a variety of flavours.   First, there is debate about whether the radio burst emission originates within or close to the magnetosphere (\citep[e.g.][]{Kumar2017}), or whether it is generated in a relativistic shock produced by an explosive energy release from the central engine (\citep[e.g.][]{Metzger2019}).  Secondly, one can consider whether the magnetar is acting in isolation, or whether its activity is stimulated by an external plasma stream (\citep[e.g.][]{Zhang2018}).  

As with pulsars, the polarimetric properties of magnetar radio bursts show diversity \citep{Kramer2007,Lower2020}.  Nonetheless, very high ($> 80$\%) linear polarisation fractions are common \citep{Magnetar_Review}, though not ubiquitous \citep{Kramer2007,Kirsten2020}.

The high linear polarisation observed for \frb\ is expected in both magnetospheric magnetar models \citep{Lu2019} and synchrotron maser shock models (\citep[e.g.][]{Metzger2019}). The magnetospheric model described in \citet{Lu2019} additionally predicts small variations of the PPA between bursts from a repeating FRB, with the burst-to-burst variability following the rotation period of the magnetar. Relativistic shock models, where the FRB emission originates much farther from the magnetar, naturally predict constant PPA within and between bursts (\citep[e.g.][]{Metzger2019}). However, small variations can be additionally explained by invoking clumpiness in the medium into which the shock front propagates, or could alternatively come from the maser emission itself (although at this time, it is unclear how large an effect this will have on the PPA).

In this work, we have observed the shortest-timescale structure seen in any FRB to date ($3-4$\,$\upmu$s), {\it and} we see that there is a range of timescales from a few $\upmu$s to a few ms. In the literature, there are bursts detected from \frb\ with total envelope widths of up to 6\,ms (at 300--800\,MHz)\cite{R3_disc,Chawla2020} (there are larger burst widths reported at 110--188\,MHz \cite{pleunis2020}, but at these low frequencies scattering dominates). It should be noted that there appears to be a frequency dependence on burst width, therefore comparison of temporal structure across different frequencies should be done with caution. The observed shortest timescales of a few $\upmu$s, and range of timescales have implications for magnetar progenitor FRB models. Assuming a magnetar progenitor, temporal fluctuations strongly constrain where the emission originates (i.e. within the magnetosphere or well outside the magnetosphere). The ratio of fluctuations to total burst duration, in our case, is approximately 4\,$\upmu$s\,/\,2\,ms\,$=$\,0.002\,$\ll$\,1, which is most naturally explained invoking emission originating within the magnetosphere \cite{beniamini2020}. The short-timescale structure observed in \frb\ implies that the emission region is on the order of 1\,km. In the case of FRB emission originating from a relativistic shock at a large distance from the magnetar, this would imply a very small area of the total shock front dominating. Nonetheless,  the temporal fluctuations could be explained by invoking clumpiness in the medium where the shock front propagates or, potentially, propagation effects. 

Additionally, we find that the bright component of B4 (B4-sb2) at 1\,$\upmu$s resolution exhibits fluctuations of $10--20\,\upmu$s. We conclude that this sub-burst is not comprised of a few well-separated bright (sub) $\upmu$s shots of emission, but it is possible that the envelope is made up of many closely spaced (sub) $\upmu$s shots of comparable amplitude. This has been predicted in models of magnetospheric burst emission, in order to explain the observed flux densities \cite{cordes2016}. 

The consistent PPA between bursts from \frb\ has direct implications regarding the precessing neutron star models\citep{zanazzi2020,levin2020} created to explain the 16.35\,day periodicity \cite{R3_period}. During precession, the line-of-sight inevitably sweeps across a much larger angular area on the neutron star surface compared to a non-precessing case. Therefore, the model not only expects PPA variation as a function of the rotational phase, it also expects PPA variation as a function of precession phase. We observe only a very small PPA variation between the bursts, which strongly suggests that the emission angle is greatly tilted from the direction of the magnetic pole in this scenario. Precession is thus disfavoured given that we require a very specific geometry to explain the constant PPA between bursts (which is also observed in the case of FRB\,20121102A \cite{Michilli2018}).

Based on the observed offset from the nearest star forming regions in the host galaxy, \citet{tendulkar2020} discuss a model in which \frb\ is a neutron star in an interacting high-mass X-ray binary system.  In this scenario, the neutron star magnetosphere is `combed' by the ionised wind of the stellar companion\citep{ioka_zhang2020}, and creates a magnetic tail as well as a clear funnel where FRBs can be seen.  Low-frequency detections of \frb\ indicate that the line-of-sight to the neutron star must still be relatively clean\citep{pleunis2020}.  Our results from microsecond polarimetry of \frb\ are consistent with the bursts being produced near the neutron star in a magnetic tail.  Nonetheless, it also remains possible that \frb's observed periodic activity is due to rotation\citep{beniamini2020}; this case can also accommodate the results we present here.

There is, arguably, a characteristic observational picture emerging for repeating FRBs (for a detailed comparison of \frb\ with other FRBs, see Methods). Specifically, repeaters exhibit the downward drifting, so-called `sad-trombone' effect\citep{Hessels2019,R3_disc}, and show narrowband burst envelopes\citep{gourdji2019,pearlman2020}. On average, repeating FRBs exhibit longer-duration burst profiles \citep{R3_disc}. Additionally, the repeaters FRB\,20121102A, \frb\ and FRB\,20190711A show remarkably consistent and characteristic polarimetric properties (highly linearly polarised, no evidence of circular polarisation, and constant PPA during and between bursts). In contrast, when including apparent non-repeaters, the global landscape of FRB polarimetric properties is diverse \citep{Caleb2018,Day2020}. As with pulsars, FRBs exhibit a wide range of polarisation fractions and PPA variations. Repeating FRBs appear to live in a very diverse set of host galaxies and local environments \citep{Chatterjee2017,Marcote2020,heintz2020}, implying that these characteristic properties are exclusive to the emission mechanism, as opposed to effects from the local medium. In this work, we have supported this characteristic observational picture of repeating FRBs with our polarisation measurements of \frb\ at 1.7\,GHz. We also suggest that the dynamic range of temporal structure of 2\,ms/4\,$\upmu$s\,$=$\,500 could be another characteristic to add to this overall description of repeating FRBs.

The results presented here highlight the importance of high-time-resolution polarimetric studies of FRBs. With lower time resolution data, narrow temporal components and subtle variations in the PPA are averaged out.  It is possible that previous flat PPA measurements from FRB\,20121102A and \frb\ are a result of this. We encourage future observations of FRBs with $\upmu$s time resolution and full polarisation information.  We also encourage searches for quasi-periodic oscillations in individual high-S/N FRBs, like the analysis conducted in this work. Studying FRBs in such fine detail is crucial for understanding their emission physics. 

\section*{Methods} 
\subsection*{Refined DM}
We refine the burst DM using the PSRCHIVE tool {\tt pdmp} to search for the DM that maxmises S/N of burst B4 (the brightest in our sample) in the 1\,$\upmu$s resolution data. The DM is found to be $348.772\pm0.006$\,$\mathrm{pc\ cm^{-3}}$, which is $0.012$\,$\mathrm{pc\ cm^{-3}}$ greater than the value the data is coherently dedispersed to. In case the burst is comprised of bright $\upmu$s shots of emission, we additionally search for the DM that maximises the peak in the profile structure (using the metric of maximising (max-min) of the time series). This was found to be $348.775$\,$\mathrm{pc\ cm^{-3}}$, consistent with the {\tt pdmp}-determined value. We thereafter incoherently dedispersed all the $1$\,$\upmu$s data to the {\tt pdmp}-determined value ($+0.012\,\mathrm{pc\ cm^{-3}}$).  This slight shift in DM does not result is significant intra-channel temporal smearing.

\subsection*{Short timescale structure in B4-sb2}
To test whether the single-time-bin spikes that appear in B4-sb2 (panel l, Figure\,\ref{fig:burstsplot_time}) are physical or consistent with amplitude-modulated noise \citep{rickett1975,cordes1976}, we remove the envelope of the burst from the data. To do this we use a Blackman window function, with a smoothing window of 19 bins, to create a model of the envelope of the burst (shown in panel c of Figure\,\ref{fig:amp_mod_noise}). This model is then divided out of the data, leaving the residuals shown in panel d of Figure\,\ref{fig:amp_mod_noise}, with off-burst noise also shown for comparison. We find no statistical outliers in this burst, implying that the narrow features are consistent with amplitude-modulated noise.

\subsection*{Power spectra modelling and statistics}
The power spectrum (in log space; see Extended Data Figure\,\ref{fig:ACFPS}) was fit with a power law of the form 
\begin{equation}\label{eq:powerlaw} f(\nu) = A\nu^{-\alpha} + C, \end{equation}
where $A$ is the amplitude, $\alpha$ is the power law slope, and $C$ is a white noise component, using the Bayesian analogue of a maximum likelihood estimation, as implemented in the Stingray modelling interface \cite{huppenkothen2019}. There are many astrophysical phenomena whose lightcurve is observed to have a power law component in the Fourier domain, often referred to as `red noise' (e.g. gamma-ray bursts; \cite{cenko2010}, active galactic nuclei; \cite{mchardy2006}, magnetars; \cite{huppenkothen2013}). We perform a goodness-of-fit test by simulating 100 fake power spectra from the best fit, and performing the same Bayesian maximum likelihood fit. The measured p-value is then the fraction of the simulations with a maximum likelihood lower than the likelihood of our fit. The results of this analysis are shown in Extended Data Table\,\ref{tab:psfit}. There are apparent oscillations in the ACF, consistent with the fluctuations seen in the profile. The power spectrum shows a power law slope, consistent with red noise. To test the statistical significance of any features in the power spectrum on top of the red noise slope, we use two metrics (for a detailed explanation, see \citet{huppenkothen2013}). First, to search for any significant narrow features in the power spectrum, we compute the residuals as a function of frequency, $\nu$, 
\begin{equation} R(\nu)=\frac{2P(\nu)}{M(\nu)},\end{equation}
where $P(\nu)$ is the power spectrum, and $M(\nu)$ is the best fit noise component. Using the Markov chain Monte Carlo (MCMC) package {\tt emcee} \citep{emcee} to generate 100 simulated residuals, we generate the distribution of max($R_{\rm sim}(\nu)$), and determine the probability that the observed peak value, max($R(\nu)$), is consistent with noise. We find no statistically significant outliers using this statistic. The second method we use is more sensitive to lower amplitude, wider features in the power spectrum, which are often referred to as quasi-periodic oscillations (QPOs) and are observed in a number of astrophysical phenomena (e.g., accreting low-mass X-ray binaries\cite{vanderklis2006}, black hole binaries\cite{remillard2006} and magnetar X-ray flares\cite{israel2005}). This second method is a model comparison method. In addition to the red noise fit described above, we fit a function with a Lorentzian describing the QPO summed with a red noise power law (as defined above). We calculate the likelihood ratio, and calibrate this likelihood ratio using MCMC simulations of the simpler model (in our case, the power law model; see \citet{protassov2002} for details). This analysis returns the posterior predictive p-value quoted in Extended Data Table\,\ref{tab:psfit}, i.e. for both B4-sb1 and B3 we cannot rule out the simpler model of a red noise power law slope. For all of the Bayesian fits described we give conservative prior distributions: flat distribution for the power law slope $\alpha$, flat distribution for the amplitude $A$, normal distribution for the white noise component $C$, and a flat distribution for the Lorentzian parameters. Since we see fluctuations of approximately 60\,$\upmu$s in the ACF of B4-sb1, we use this as the inital guess for the centroid frequency of the Lorentzian.

The bright envelope of burst B4-sb2 dominates in both the ACF and power spectrum, and so any features associated with quasi-periodic oscillation are difficult to detect. One way to bypass this issue would be to remove the envelope (divide out a smooth model of the burst envelope), but this can introduce features in the power spectrum which are not physical \cite{huppenkothen2013}. We therefore only perform this analysis on burst B3 and B4-sb1, which do not have a prominent envelope that would dominate the results. The results are shown in Extended Data Figure\,\ref{fig:ACFPS} and Extended Data Table\,\ref{tab:psfit}. 

\subsection*{Polarimetric calibration and RM measurement}
We did not perform an independent polarisation calibrator scan to use for polarimetric calibration. Instead, we use the test pulsar observation of \psr\ to determine the calibration solutions to apply to our target data. A similar polarimetric calibration technique was used for radio bursts detected from SGR~1935$+$2154 using voltage data with the VLBI backend of the Westerbork single-dish telescope RT1 \citep{Kirsten2020}.

We assume that any leakage between the two polarisation hands only affects Stokes~V (defined as V = LL $-$ RR using the PSR/IEEE convention \cite{vanstraten2010}). We also assume that the delay between the two polarisation hands only significantly affects Stokes~Q and U. The calibration we apply ignores second-order effects. We performed a brute force search for the RM that maximises the linear polarisation fraction using the PSRCHIVE tool {\tt rmfit}. With {\tt rmfit} we select a range of RMs to search, in a number of trial steps. The delay between the polarisation hands approximately manifests as an offset from the true RM of the source, assuming the delay is frequency-independent. For this reason, we select a larger range of RMs than what would be motivated by the known measured RM of \frb\ ($- 114.6\,\mathrm{rad\ m^{-2}}$)\citep{R3_disc}, and we search from $-$5000$\,\mathrm{rad\ m^{-2}}$ to 5000$\,\mathrm{rad\ m^{-2}}$ in 500 equally spaced steps.

For \psr, we measure an RM of $-657\,\mathrm{rad\ m^{-2}}$, which is $438$ units from the true RM of \psr\ ($-218.7\,\mathrm{rad\ m^{-2}}$) \cite{force2015}. This approximately translates to a delay of 5.5$\,\mathrm{ns}$. We use the {\tt rmfit}-RM to Faraday correct the pulsar data, and we reproduce the polarimetric properties and PPA of \psr\ within 8\% of published properties \citep{gould1998}. Extended Data Figure\,\ref{fig:pulsar} illustrates the calibration we applied. We note that we had $<1\,\mathrm{minute}$ on \psr, which has a rotational period of approximately 1$\,\mathrm{s}$ \citep{arzoumanian1994}, so it is likely that our observed average profile did not completely stabilise to the published average profile, which is based on the sum of many more individual pulses. As such, the aforementioned 8\% deviation should be treated as an upper limit on the inaccuracy of the polarimetric calibration. 

We assume there are no significant changes to the calibration required between the test pulsar scan and the detected \frb\ bursts ($<$\,1\,$\mathrm{hr}$ between the \psr\ scan and burst B1). Bursts B1 and B4 have a sufficient S/N to determine an RM using {\tt rmfit} (S/N values determined in \citet{Marcote2020} are quoted in Extended Data Table\,\ref{tab:pol_properties}). B1 and B4 are separated in time by $>$\,4\,$\mathrm{hr}$ and we note their measured {\tt rmfit}-RMs differ by approximately $8$ units (1--2\% of the measured value).  Thus we conclude that the bursts have consistent RMs. The measured {\tt rmfit}-RM for B4 is $-536\,\pm\,5\,\mathrm{rad\ m^{-2}}$, which when combined with the offset due to a delay between the polarisation hands ($+438$ units, measured using the \psr\ data) gives a true RM of $-98\,\mathrm{rad\ m^{-2}}$. We note that we have not removed the delay between polarisation hands from the data before running {\tt rmfit}, and thus the error quoted is lower than the true error as it does not reflect uncertainties associated with the covariance between RM and delay. This is tackled in the following steps using a Q-U fit. 

To better determine the burst RM and associated errors, we perform a joint least squares fit of Stokes~Q and U spectra (as a function of frequency, $\nu$), using the following equations:
\begin{equation} Q/I = L\cos(2(c^2\mathrm{RM}/\nu^2 + \nu \pi D + \phi)),\end{equation}
\begin{equation} U/I = L\sin(2(c^2\mathrm{RM}/\nu^2 + \nu \pi D + \phi)),\end{equation}
where $c$ is the speed of light, and the free parameters $L$, the linear polarisation fraction, $D$, the delay between polarisation hands, and $\phi=\phi_{\infty}+\phi_{\rm inst}$, where $\phi_{\infty}$ is the absolute angle of the polarisation on the sky (referenced to infinite frequency), and $\phi_{\rm inst}$ is the phase difference between the polarisation hands. We perform the joint fit on Q/I and U/I spectra for \psr\ and for burst B4, where the delay, $D$, is assumed to be the same for both the pulsar and target scans. We fix the known RM of \psr \citep{force2015}, $-218.7\,\mathrm{rad\ m^{-2}}$. We measure $D=5.4\pm0.2\,\mathrm{ns}$, consistent with our prediction from the offset in RM from the true RM of \psr\ using {\tt rmfit}. Additionally, we measure the RM of burst B4 to be $-104\pm 20\,\mathrm{rad\ m^{-2}}$, where the large fractional error arises due to covariances between the fit parameters (RM, $D$ and $\phi$) that could not be removed as we did not record independent information from a polarisation calibrator source. We find the RM to be consistent with the previously measured RM values for \frb\ \citep{R3_disc,Chawla2020}.

To correct for parallactic angle, we rotate the linear polarisation vector by
\begin{equation} \theta = 2 \tan^{-1}\left({\frac{\sin({\rm HA})\cos(\phi)}{(\sin(\phi)\cos(\delta)-\cos(\phi)\sin(\delta)\cos({\rm HA})}}\right),\end{equation}
where ${\rm HA}$ is the hour angle of the burst, $\phi$ is the latitude of Effelsberg, and $\delta$ is the declination of \frb. The parallactic angle corrected PPA is shown in the top panel of each sub-figure in Figure~\ref{fig:burstsplot}. We plot the probability distribution of PPA per time bin, following \citet{everett2001}, and mask any bins where the unbiased linear S/N is below 3.

\subsection*{Is there a characteristic observational description of repeating FRBs?}

Our high-time-resolution, polarimetric measurements of \frb\ demonstrate remarkable phenomenological similarity to FRB\,20121102A \citep{Michilli2018,Gajjar2018}.  Both of these repeating FRBs show 20-30\,$\upmu$s sub-bursts (in some high-S/N bursts, at least), approximately 100\% linear polarisation, approximately 0\% circular polarisation, and a constant PPA during the bursts. Moreover, between 16 bursts found in three observations spanning 25 days, \citet{Michilli2018} report consistent PPAs throughout. \citet{Michilli2018} fit for a variable RM per day, but a global PPA for all epochs.  \citet{Gajjar2018} quote different average PPAs between bursts, but this is potentially because they allow the RM to vary between bursts detected within approximately 1 hour.  The covariance between RM and PPA makes it difficult to distinguish small variations in the former compared to the latter.  Here we find that the PPA of \frb\ is also remarkably similar {\it between} bursts, as shown in Figure~\ref{fig:burstsplot} and Extended Data Table\,\ref{tab:pol_properties}.

Comparing our 1.7-GHz measurements with the available $110-188$\,MHz \citep[LOFAR][]{pleunis2020}, $300$--$400$\,MHz \citep[GBT][]{Chawla2020} and $400$--$800$\,MHz \citep[CHIME/FRB][]{R3_disc} bursts, we find that the polarimetric properties are also persistent over at least four octaves in radio frequency.  However, the lack of absolute PPA calibration prevents us from investigating whether the average PPA is both persistent in time and between radio frequencies.  In the case of FRB\,20121102A, it appears that the linear polarisation fraction decreases towards lower frequencies (Plavin et al., in prep.).  It is, as yet, unclear whether that is due to an intrinsic change in the emission physics, or whether it reflects a propagation effect.  The RM of FRB\,20121102A is highly variable \citep[][]{Michilli2018,Gajjar2018,hilmarsson2020}, and 2$-$3 orders of magnitude larger than \frb.  The association of FRB\,20121102A with a persistent, compact radio source \citep{Chatterjee2017,Marcote2017} -- whereas none is detected coincident with \frb\ \citep{Marcote2020} -- further demonstrates that their local environments are different, despite both being near to a star-forming region \citep{Tendulkar2017,Bassa2017,Marcote2020,tendulkar2020}.

Regardless of differences in host galaxy type and the local environment, however, the remarkable similarity of burst properties demonstrates that FRB\,20121102A and \frb\ have the same physical origin.  This is further emphasised by the detection of periodicity in the burst activity rate of \frb\citep{R3_period} with $P_{\rm activity} \sim$16\,day, and the potential detection of a similar effect from FRB\,20121102A\citep{Rajwade2020,Cruces2020} with $P_{\rm activity}\sim$157\,day.

To date, the only other repeating FRB that has polarisation information from more than one burst, and is localised to a host galaxy, is FRB\,20190711A \citep{Day2020,Kumar2020}. FRB\,20190711A clearly shows the downward-drifting `sad trombone' effect characteristic of repeating FRBs \citep{Hessels2019,R2_disc}. Also, the polarimetric properties of FRB\,20190711A show a striking observational similarity; it is also highly linearly polarised, approximately 0\% circularly polarised and has a constant PPA across the burst profiles. FRB\,20190711A has been localised to a star-forming galaxy \citep{macquart2020,heintz2020}, different from the hosts of FRB\,20121102A (found in a faint starburst galaxy \cite{Chatterjee2017}) and \frb\ (localised to a massive quiescent galaxy \cite{Marcote2020}).

Recently, \citet{Luo2020} report the polarisation properties of $7$ bursts from the repeating FRB\,20180301A. FRB\,20180301A shares a number of properties with other well-studied repeating FRBs, including downward-drifting sub-bursts, narrowbandedness,  and no evidence of circular polarisation. However, the approximately $100\,\%$ linear polarisation and flat PPA across burst profiles is not always observed in the case of FRB\,20180301A \cite{Luo2020}.

The so-far non-repeating FRB\,20181112A shows 4 sub-bursts spanning a total burst duration of 1.5\,ms, with different apparent RMs and DMs between sub-bursts \citep{Cho2020}.  \citet{Day2020} also found similar effects in their sample of five FRBs.  The apparent RM variations of approximately 10--20\,rad\,m$^{-2}$ seen in the ASKAP FRB sample are too subtle to probe for \frb\ given the data we present here and the uncertainty on the delay calibration.  We note, however, that (apparent) RM variations at this level are likely excluded based on previously published \frb\ polarimetric results taken at $110-188$\,MHz with LOFAR\citep{pleunis2020}, $300-400$\,MHz using GBT\citep{Chawla2020} and $400-800$\,MHz using CHIME/FRB\citep{R3_disc} because they would lead to a lower polarisation fraction than observed.

Nonetheless, at the high-time-resolution afforded by these data, we detect subtle PPA variations of a few degrees between sub-bursts lasting $\lesssim 100$\,$\upmu$s each.  This is most visible for burst B1 (Figure~\ref{fig:burstsplot}).  For the bright, 60\,$\upmu$s dominant component of B4 (B4-sb2), where we have maximum S/N per unit time, there is the suggestion of PPA variations of a few degrees, when studying this component at $1\,\mathrm{\upmu s}$ time resolution (Figure \ref{fig:burstsplot}). This could be interpreted as potential small PPA swings, or that this burst component is actually composed of many sub-$\mathrm{\upmu s}$ components with PPAs that vary on the level of a few degrees, similar to what we see between the $100$\,$\mathrm{\upmu s}$ burst components.

\frb\ shows some of the shortest-timescale temporal features seen in any FRB to date. For comparison, FRB\,20121102A, FRB\,20170827A and FRB\,20181112A have shown 30\,$\upmu$s substructures \citep{Michilli2018,Farah2018,Cho2020}. In the case of FRB\,20170827, the burst shows a single component of width $30$\,$\upmu$s \citep{Farah2018}, and similarly, FRB\,20121102A produced a single burst of width $30$\,$\upmu$s \citep{Michilli2018}. FRB\,20181112A, also shows a single narrow component, but the results are limited by scattering at the 20\,$\upmu$s level \citep{Cho2020}. In this work, we have demonstrated that not only does \frb\, also exhibit short-duration components similar to what has been seen in other FRBs (e.g. the $30$\,$\upmu$s spike in the inset on Figure\,\ref{fig:burstsplot}), but, in fact, we observe temporal scales spanning three orders of magnitude, the shortest reaching only a few $\upmu$s. 

\citet{Marcote2020} estimated a Galactic scattering time of $2.7$\,$\upmu$s at 1.7\,GHz from the measurement of the scintillation bandwidth. Independently, \citet{Chawla2020} place a constraint on the scattering timescale of \frb\ of $\tau<$\,1.7\,ms at $350$\,MHz, which, assuming a frequency scaling of $\tau\propto\nu^{-4}$, gives a scattering time at $1.7$\,GHz of $<$\,3\,$\upmu$s, consistent with \citet{Marcote2020}. The shortest timescale structure observed in this work is consistent with this scattering prediction. Our results rule out that burst B4 is composed of a few extremely bright sub-$\upmu$s shots of emission well-spaced in time, similar to what is observed in Crab giant pulses \citep{Hankins2003}. If the $20$\,$\upmu$s morphology that we observe in the profile of B4 are made up of sub-$\upmu$s shots of emission, they must be closely packed in time and of approximately equal amplitude.

\section*{Data availability} 
The data that support the plots and results in this study are available from \url{https://doi.org/10.5281/zenodo.4350456}, or from the corresponding author upon reasonable request.

\section*{Code availability} 
The code used to analyse the data and create the figures in this work can be found here: \url{https://github.com/KenzieNimmo/Microsecond_Polarimetry_R3}


\section*{Additional information}
Correspondence and requests for material should be addressed to K.N.
\section*{Acknowledgements}
{
We thank Pawan Kumar, Brian Metzger, Lorenzo Sironi, Maxim Lyutikov, Michiel van der Klis and Phil Uttley for helpful discussions.

The European VLBI Network is a joint facility of independent European, African, Asian, and North American radio astronomy institutes. Scientific results from data presented in this publication are derived from the following EVN project code: EM135.
This work was also based on simultaneous EVN and PSRIX data recording observations with the 100-m telescope of the MPIfR (Max-Planck-Institut f\"{u}r Radioastronomie) at Effelsberg, and we thank the local staff for this arrangement.

J.W.T.H. acknowledges funding from an NWO Vici grant (“AstroFlash”; VI.C.192.045).
F.K. acknowledges support by the Swedish Research Council.
B.M. acknowledges support from the Spanish Ministerio de Econom\'ia y Competitividad (MINECO) under grant AYA2016-76012-C3-1-P and from the Spanish Ministerio de Ciencia e Innovaci\'on under grants PID2019-105510GB-C31 and CEX2019-000918-M of ICCUB (Unidad de Excelencia ``Mar\'ia de Maeztu'' 2020-2023).}

\section*{Author contributions}

K.N. discovered the signals, led the data analysis, made the figures, and wrote the majority of the manuscript.  J.W.T.H. guided the work and made important contributions to the writing and interpretation.  A.K. performed pre-processing of the EVN voltage data.  All other authors contributed significantly to aspects of the data acquisition, analysis strategy or interpretation.

\section*{Competing interests}
The authors declare no competing interests.

\clearpage
\newpage
\section*{Supplementary Information}
\onecolumn
\setcounter{figure}{0}
\captionsetup[figure]{name={\bf Extended Data Figure}}

\setcounter{table}{0}
\captionsetup[table]{name={\bf Extended Data Table}}

\begin{figure*}
\resizebox{\hsize}{!}
        {\includegraphics[trim=0cm 0.6cm 0cm 0.9cm, clip=true,width=\textwidth,height=195mm]{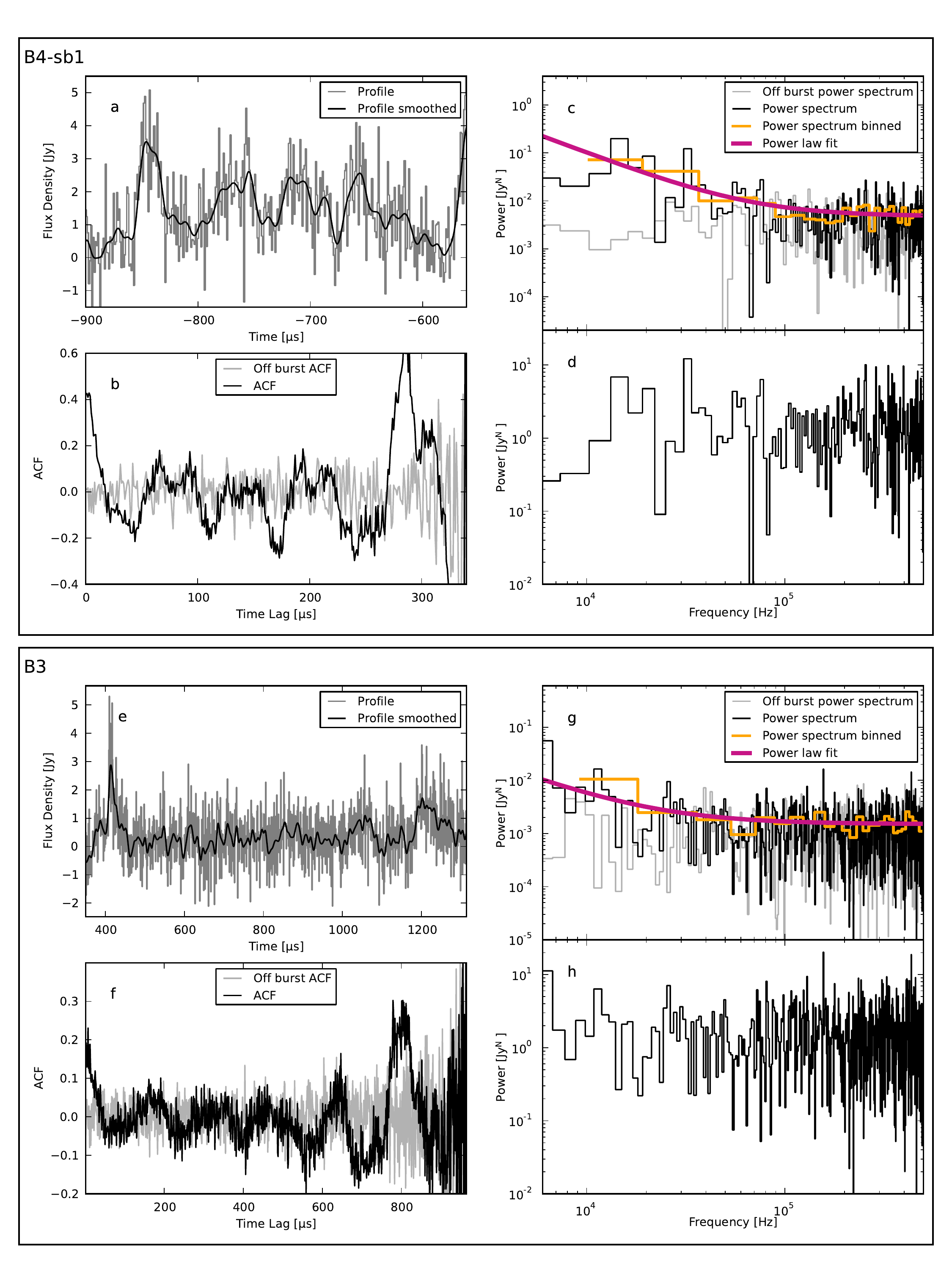}}
  \caption{The temporal profiles of bursts B4-sb1 (top frame) and B3 (bottom frame) are shown in the top left of each frame (panels a and e). The corresponding autocorrelation functions (ACF) of these temporal profiles are shown in panels b and f. The corresponding power spectra of the temporal profiles are shown in panels c and g, and the orange line is the power spectrum downsampled in frequency by a factor of 3. Overplotted in pink on the power spectrum is the best fit power law plus white noise component. Panels d and h are the residuals (2$\times$power spectrum/best fit model). }
     \label{fig:ACFPS}
\end{figure*}

\begin{table*} 
\caption{\label{tab:psfit}Results of power spectrum fitting}
\resizebox{\textwidth}{!}{\begin{tabular}{lcccc}
\hline
\hline
Burst & Power law slope ($\alpha$) & Goodness of fit p-value$^{\rm a}$ & Posterior predictive p-value$^{\rm b}$ & Residual outlier p-value$^{\rm c}$\\
\hline
B4-sb2 & 1.58\,$\pm$\,0.02 & 0.44 & 0.44 & 0.37\\
B3 & 1.38\,$\pm$\,0.01 & 0.33 &  0.74 & 0.06 \\
\hline
\multicolumn{5}{l}{$\mathrm{^{a}}$Goodness of fit of the power law red noise model descibed by Equation \ref{eq:powerlaw}.}  \\
\multicolumn{5}{l}{$\mathrm{^{b}}$Model comparison of power law red noise model, versus a power law plus Lorentzian (to describe a quasi-periodic oscillation). See text for details.}  \\
\multicolumn{5}{l}{$\mathrm{^{c}}$p-value of the highest outlier in the residuals of the power spectrum divided by the best fit power law slope.}  \\
\end{tabular}}
\end{table*}

\begin{table*}
\caption{\label{tab:pol_properties}Burst polarisation properties and polarisation position angle fit results.}
\resizebox{\textwidth}{!}{\begin{tabular}{lcccccccc}
\hline
\hline
		{Burst}& {MJD$^{\rm a}$}   & {Fluence [Jy ms]$^{\rm a,b}$} & {S/N ${^{\rm a}}$} & {L$_{\rm unbiased}$/I [\%]$^{\rm c,d,e}$} & {V/I [\%]$^{\rm d,e}$} & {PPA offset [deg]$^{\rm d}$} & $\chi^{2}$ & Degrees of freedom\\
		\hline
		B1 &  58653.0961366466  & 0.72 & 9.87 & 112 $\pm$ 14  & $-$1 $\pm$ 12  & $-$2.96 & 49.7 & 25\\
        B2 &  58653.1112573504 & 0.20 & 9.61 & 88 $\pm$ 20 & $-$4 $\pm$ 20 & $-$5.85 & 0.91 & 2\\
        B3 &  58653.1465969404 & 0.62 & 9.78 & 99 $\pm$ 14  & $-$15 $\pm$ 14 & 0.30 & 25.55 & 14\\
        B4 & 58653.2785078914 & 2.53 & 65.42 & 103 $\pm$ 4 & 5 $\pm$ 4 & 0.02 & 92.83 & 81\\
        B4-sb2 (1\,$\upmu$s) & - & - & - & - & - & $-$0.50 & 445.77 & 121\\
		\hline
		\multicolumn{8}{l}{$\mathrm{^{a}}$For details on the determination of these values see \citet{Marcote2020}.}  \\
		\multicolumn{8}{l}{$\mathrm{^{b}}$A conservative fractional error of $30\%$ is taken for the derived fluences.} \\
		\multicolumn{8}{l}{$\mathrm{^{c}}$Removing the baseline can result in the condition ${\rm I}^2 \ge {\rm Q}^2 + {\rm U}^2 + {\rm V}^2$ not being satisified, }\\
		\multicolumn{8}{l}{\hspace{0.3cm} which can lead to apparent linear polarisation fractions $> 100\%$.}\\
		\multicolumn{8}{l}{$\mathrm{^{d}}$The fractional polarisations and PPA values are measured over the Gaussian-fit 2-$\sigma$ region of the burst profile,}\\
        \multicolumn{8}{l}{$^{\rm e}$The quoted uncertainties are statistical $2$-$\sigma$ errors assuming the errors in the Stokes parameters are independent of each other,}\\
         \multicolumn{8}{l}{\hspace{0.3cm} and the errors in the time bins are independent of each other.}\\
          \multicolumn{8}{l}{\hspace{0.3cm} The uncertainties do not contain the calibration uncertainty nor do they encapsulate the effect of removing the baseline. }\\
        
\end{tabular}}
        
\end{table*}

\begin{figure*}
\resizebox{\hsize}{!}
        {\includegraphics{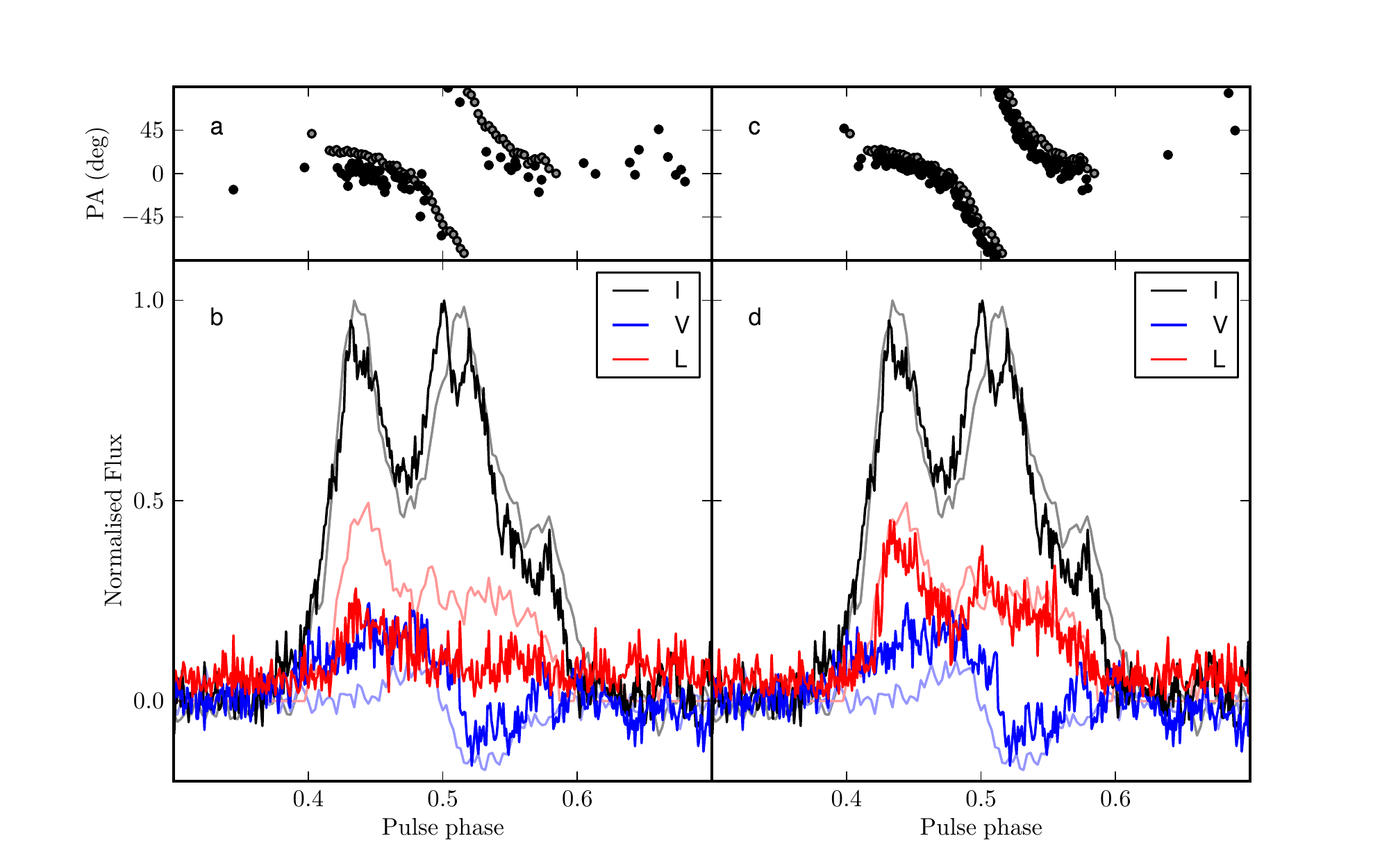}}
  \caption{The average polarisation profiles (panels b and d) and polarisation position angle (panels a and c) of \psr. Black represents the Stokes I profile, red is the unbiased linear polarisation profile (defined in \citet{everett2001}, and rewritten here in Equation 1), and blue is the circular polarisation (Stokes V) profile. Panels a and b show the polarisation profile and position angle after Faraday-correcting to the true rotation measure\citep{force2015} of \psr\ ($-218.7\,\mathrm{rad\ m^{-2}}$); here we are not correcting for the instrumental delay between polarisation hands. Panels c and d are Faraday-corrected with the rotation measure determined using the PSRCHIVE tool {\tt rmfit}, which, in essence, accounts for the instrumental delay. For comparison, we plot the profile and position angle from the literature using more transparent colours \citep{gould1998}. This illustrates the calibration we applied to the bursts from \frb.}
     \label{fig:pulsar}
\end{figure*}

\end{document}